\documentclass[11pt]{article}

\usepackage{amsfonts,amsthm,amsmath,amssymb,bbm,mathrsfs,verbatim,paralist,mathtools}
\usepackage{epsfig,color}
\usepackage{xcolor}
\usepackage[sort&compress,numbers]{natbib}
\usepackage{graphicx}
\usepackage[font=small,labelfont=bf]{caption}

\numberwithin{equation}{section}

\newcommand{\be}{\begin{equation}}
\newcommand{\ee}{\end{equation}}
\newcommand{\bea}{\begin{align}}
\newcommand{\eea}{\end{align}}

\newcommand{\abs}[1]{\lvert#1\rvert}

\def\rw{\rightarrow}

\makeatletter
\def\underbracket{\@ifnextchar [
  {\@underbracket} {\@underbracket [\@bracketheight]}}
\def\@underbracket[#1]{\@ifnextchar [
  {\@under@bracket[#1]} {\@under@bracket[#1][0.2em]}}
\def\@under@bracket[#1][#2]#3{
  \mathop {\vtop {\m@th \ialign {##
           \crcr $\hfil \displaystyle {#3}\hfil$
           \crcr \noalign {\kern 3\p@ \nointerlineskip }\upbracketfill 
{#1}{#2}
           \crcr \noalign {\kern 3\p@ }
           \crcr }}}\limits}
\def\upbracketfill#1#2{$\m@th \setbox \z@ \hbox {$\braceld$}
                 \edef\@bracketheight{0.5pt}
                 \upbracketend{#1}{#2}
                 \leaders \vrule \@height #1 \@depth \z@ \hfill
                 \leaders \vrule \@height #1 \@depth \z@ \hfill
                 \upbracketend{#1}{#2}$}
\def\upbracketend#1#2{\vrule height #2 width #1\relax}

\def\overbracket{\@ifnextchar [
{\@overbracket} {\@overbracket [\@bracketheight]}}
\def\@overbracket[#1]{\@ifnextchar [
{\@over@bracket[#1]} {\@over@bracket[#1][0.2em]}}
\def\@over@bracket[#1][#2]#3{
\mathop {\vbox {\m@th \ialign {##
 \crcr \noalign {\kern 3\p@ }
 \downbracketfill {#1}{#2}
 \crcr \noalign {\kern 3\p@ \nointerlineskip }
 \crcr  $\hfil \displaystyle {#3}$
 \crcr
 } }}\limits}
\def\downbracketfill#1#2{$\m@th
 \setbox \z@ \vbox {$\braceld$}
 \edef\@bracketheight{0.5pt}
 \downbracketend{#1}{#2}
 \leaders \vrule \@height #1 \@depth \z@ \hfill
 \leaders \vrule \@height #1 \@depth \z@ \hfill
\downbracketend{#1}{#2}$}
\def\downbracketend#1#2{\vrule depth #2 width #1\relax}
\makeatother

\def\ce{competition effect}
\def\dn{distortion}

\newcommand{\Comp}{{\mathcal C}}
\newcommand{\Dist}{{\mathcal D}}

\title{The energy costs of biological insulators}
\author{{John Barton$^{1,2,\footnote{Work done at: Department of Physics, Rutgers University, Piscataway, NJ 08854 USA}}$ and Eduardo D. Sontag$^3$}\\ \\
{\small $^1$ Department of Chemical Engineering, MIT,}\\
{\small Cambridge, MA 02139 USA}\\ 
{\small $^2$ Ragon Institute of Massachusetts General Hospital,}\\
{\small MIT, and Harvard University, Boston, MA 02129 USA} \\
{\small $^2$ Department of Mathematics, Rutgers University,}\\
{\small Piscataway, NJ 08854-8019 USA}
}

\begin{document}

\maketitle

\begin{abstract}

Biochemical signaling pathways can be insulated from impedance and competition
effects through enzymatic ``futile cycles'' which consume energy, typically in
the form of ATP.  We hypothesize that better insulation necessarily requires
higher energy consumption, and provide evidence, through the computational
analysis of a simplified physical model, to support this hypothesis.

\end{abstract}

\section{Introduction} \label{introduction}

An important theme in the current molecular biology literature is the
attempt to understand cell behavior in terms of cascades and feedback
interconnections of more elementary ``modules,'' which may be re-used in
different pathways~\citep{lauffenburgermodules,modules}.
Modular thinking plays a fundamental role in the prediction of the behavior of a
system from the behavior of its components, guaranteeing that the properties
of individual components do not change upon interconnection.
Intracellular signal transduction networks are often thought of as modular
interconnections, passing along information while also amplifying or
performing other signal-processing tasks.  It is assumed that their operation
does not depend upon the presence or absence of downstream targets to which
they convey information.  However, just as electrical, hydraulic, and other
physical systems often do not display true modularity, one may expect that
biochemical systems, and specifically, intracellular protein signaling
pathways and genetic networks, do not always ``connect'' in an ideal modular
fashion.

Motivated by this observation, the paper~\citep{DelVecchio:2008gy} dealt with
a systematic study of the effect of interconnections on the input/output
dynamic characteristics of signaling cascades. 
Following~\citep{saez}, the term \emph{retroactivity} was
introduced in order to generically refer to such effects, which constitute an
analog of non-zero output impedance in electrical and mechanical systems, and
retroactivity in several simple models was quantified.  It was shown how
downstream targets of a signaling system (``loads'') can produce changes in
signaling, thus ``propagating backwards'' (and ``sideways'') information about
targets.  Further theoretical work along these lines was reported
in~\citep{ddv_ecc09,mra_book10}.  Experimental verifications were reported
in~\cite{ventura2010} and in~\citep{Jiang:2011ch}, using a covalent modification
cycle based on a reconstituted uridylyltransferase/uridylyl-removing enzyme
PII cycle, which is a model system derived from the nitrogen assimilation
control network of \emph{Escherichia coli}.

The key reason for retroactivity is that signal transmission in biological
systems involves chemical reactions between signaling molecules.
These reactions take a finite time to occur, and during the process, while
reactants are bound together, they generally cannot take part in the other
dynamical processes that they would typically be involved in when unbound.
One consequence of this ``sequestering'' effect is that influences are also
indirectly transmitted ``laterally,'' in that for a single input - multiple
output system, the output to a given downstream system is influenced by other
outputs.

In order to attenuate the effect of retroactivity, the
paper~\citep{DelVecchio:2008gy} proposed a negative feedback mechanism
inspired by the design of operational amplifiers (``OpAmps'') in electronics,
employing a mechanism implemented through a covalent modification cycle based on
phosphorylation/dephosphorylation reactions.  For appropriate parameter
ranges, this mechanism enjoys a remarkable insulation property, having an an
inherent capacity to shield upstream components from the influence of 
downstream systems and hence to increase the modularity
of the system in which it is placed.  One may speculate whether this is indeed
one reason that such mechanisms are so ubiquitous in cell signaling pathways.
Leaving aside speculation, however, one major potential disadvantage of
insulating systems based on ``OpAmp'' ideas is that they impose a metabolic
load, ultimately because amplification requires energy expenditure.

Thus, a natural question to ask from a purely physical standpoint is:
\emph{does better insulation require more energy consumption?}  This is the
subject of the present work.  We provide a qualified positive answer: for a
specific (but generic) model of covalent cycles, natural notions of insulation
and energy consumption, and a Pareto-like view of multi-objective optimization,
we find, using a numerical parameter sweep in our models, that better
insulation indeed requires more energy.

In addition to this positive answer in itself, two major contributions of this
work are:
\begin{itemize}
\item[(a)]
we introduce two innovative measures of retroactivity, and dually insulation,
in terms of two competing goals: (1) the minimization of the difference
between the output of the insulator and the ``ideal'' behavior,
and (2) the attenuation of the {\ce}, the change in output when a new downstream
target is added;
\item[(b)]
we introduce a new way to characterize optimality through balancing of these
goals in a Pareto sense.
\end{itemize}
These contributions should be of interest even in other studies of insulation
that do not involve energy use. 

We remark that the recent paper by Lan et al.~\citep{Lan:2012natphys} dealt with
the need for energy dissipation when improving adaptation speed and
accuracy, in the context of perfect adaptation.  Although we also ask about
energy costs, our technical quantification of energy use, and also the problem
that we consider, are very different.
Also somewhat related is the recent paper by Shoval et
al.~\citep{Shoval:2012ke}, which also considers Pareto optimality in a
biological context of balancing competing phenotype objectives.  Again, our
work is different, since a very different type of problem is analyzed.

\section{Statement of the problem}

We are interested in biological pathways which transmit a single,
time-dependent input signal to one or more downstream targets. A prototypical
example is a transcription factor Z which regulates the production of one or
more proteins by binding directly to their promoters, forming a
protein-promoter complex. Assuming a single promoter target for simplicity,
this system is represented by the set of reactions
\begin{align} \label{DC reactions}
\begin{aligned}
\emptyset \overset{k(t)}{\rw} \mbox{Z} \overset{\delta}{\rw} \emptyset, \\
\mbox{Z+p} \overset{k_{\rm on}}{\underset{k_{\rm off}}{\rightleftharpoons}} \mbox{C},
\end{aligned}
\end{align}
where p stands for the promoter and C denotes the protein-promoter complex. 
An analogous set of chemical reactions can be used to describe a signaling
system in which Z denotes the active form of a kinase, and p is a protein
target, which can be reversibly phosphorylated to give a modified form C. For
our analysis, the particular interpretation of Z, p, and C will not be
important.
One thinks of Z as describing an ``upstream'' system that regulates the
``downstream'' target C.  Although mathematically the distinction between
``upstream'' and ``downstream'' is somewhat artificial, the roles of
transcription factors as controllers of gene expression, or of enzymes on
substrate conversions, and not the converse, are biologically natural and
accepted.

\subsubsection*{Mathematical model of the basic reaction}

We adopt the convention that the (generally time-dependent) concentration of
each species is denoted by the respective italics symbol; for example,
$X=X(t)$ is the concentration of X at time $t$.
We assume that the transcription factor Z is produced or otherwise activated
at a time-dependent rate $k(t)$, and decays at a rate proportional to a
constant $\delta$, and that the total concentration of the promoter 
$p_{\rm tot}$ is fixed. This leads to a set of ordinary differential equations
(ODEs) describing the dynamics of the system,
\begin{align} \label{DC ODEs}
\begin{aligned}
\frac{d Z}{d t} &= k(t)-\delta Z - k_{\rm on} Z \left(p_{\rm tot} - C\right) + k_{\rm off} \,C, \\
\frac{d C}{d t} &= k_{\rm on} \left(p_{\rm tot} - C\right)Z - k_{\rm off} \, C.
\end{aligned}
\end{align} 
The generalization of these equations to the case of multiple output targets
is straightforward. Protein synthesis and degradation take place on time
scales that are orders of magnitude larger than the typical time scales of
small molecules binding to proteins, or of transcription factors binding to
DNA~\citep{alonbook}.  Thus we will take the rates $k(t)$ and $\delta$ to be
much smaller than other interaction rates such as $k_{\rm on}$ and 
$k_{\rm off}$. In \eqref{DC reactions} $Z$ represents the input, and $C$ the
output.

\subsubsection*{The ideal system and the {\dn} measure}

Sequestration of the input Z by its target p affects the dynamics of the
system as a whole, distorting the output to C as well as to other potential
downstream targets.  In an ``ideal'' version of \eqref{DC ODEs}, where
sequestration effects could be ignored, the dynamics would instead be given by 
\begin{align} \label{ideal ODEs}
\begin{aligned}
\frac{d Z}{d t} &= k(t)-\delta Z, \\
\frac{d C}{d t} &= k_{\rm on} \left(p_{\rm tot} - C\right)Z - k_{\rm off} \, C.
\end{aligned}
\end{align} 
The term $- k_{\rm on} Z \left(p_{\rm tot} - C\right) + k_{\rm off}\, C$ that
was removed from the first equation represents a \emph{retroactivity} term, in
the language of~\citep{DelVecchio:2008gy}.  This is the term that quantifies
how the dynamics of the ``upstream'' species Z is affected by its
``downstream'' target C.
In the ``ideal'' system~\eqref{ideal ODEs}, the transmission of the signal from
input to output is undisturbed by retroactivity effects. 
We thus use the (relative) difference between the output signal in a system
with realistic dynamics and the ideal output, as given by the solution of
\eqref{ideal ODEs}, as a measure of the output signal distortion, and
define the \emph{\dn} $\Dist$ to be
\begin{equation} \label{DN}
\Dist = \frac{1}{\sigma_{C_{\rm ideal}}} \, \langle \abs{C_{\rm ideal}(t) - C_{\rm real}(t)} \rangle,
\end{equation} 
where $\langle \cdot \rangle$ denotes a long time average. 
Here we normalize by dividing by the standard deviation of the ideal signal,
\begin{equation} \label{sigma Cideal}
\sigma_{C_{\rm ideal}} = \sqrt{\langle (C_{\rm ideal}(t)-\langle C_{\rm ideal}(t)\rangle)^2 \rangle} .
\end{equation} 
Thus \eqref{DN} measures the difference between the output in the real and ideal systems,
in units of the typical size of the time-dependent fluctuations in the ideal output signal.

\subsubsection*{Fan-out: multiple targets}

Another consequence of sequestration effects is the interdependence of the
output signals to different downstream targets connected in parallel.  Each
molecule of Z may only bind to a single promoter at a time, thus introducing a
competition between the promoters to bind with the limited amount of Z in the
system.
This is a question of practical interest, as transcription factors typically
control a large number of target genes.  For example, the tumor suppressor
protein p53 has well over a hundred targets~\cite{riley07}.  
A similar issue appears in biochemistry, where promiscuous enzymes may affect
even hundreds of substrates.  For example, alcohol dehydrogenases (ADH)
target about a hundred different substrates to break down toxic alcohols and
to generate useful aldehyde, ketone, or alcohol groups during biosynthesis of
various metabolites \cite{adolph2000structural}. 

We quantify the size of this {\ce} by the change in an output signal to a
given target in response to an infinitesimal change in the 
abundance
of another parallel target. For definiteness, consider~\eqref{DC reactions}
with an additional promoter p$^\prime$, which bonds to Z to form a complex
C$^\prime$ with the same on/off rates as p:
\[
\mbox{Z+p}^\prime \overset{k_{\rm on}}{\underset{k_{\rm off}}{\rightleftharpoons}} \mbox{C}^\prime\,,
\]
and the corresponding equation added to \eqref{DC ODEs},
\[
\frac{d C'}{d t} = k_{\rm on} \left(p'_{\rm tot} - C'\right)Z - k_{\rm off} \, C'.
\]
We then define the \emph{\ce} of the system \eqref{DC ODEs} as 
\begin{equation} \label{CE}
\Comp = \frac{1}{\sigma_C}\,\left\langle \left. \left(\frac{\partial \, C(t)}{\partial p^\prime_{\rm tot}}\right) \right|_{p^\prime_{\rm tot}=0} \right\rangle .
\end{equation} 
Again we normalize by the standard deviation of the output signal $C(t)$,
\begin{equation} \label{sigma C}
\sigma_{C} = \sqrt{\langle (C(t)-\langle C(t)\rangle)^2 \rangle} ,
\end{equation} 
computed with $p'_{\rm tot}=0$, so that \eqref{CE} measures the change in the 
output signal when an additional target is introduced relative to the size of the
fluctuations of the output in the unperturbed system.

\subsubsection*{Design of an insulator}

\begin{figure}
\centering
\includegraphics[scale=.43]{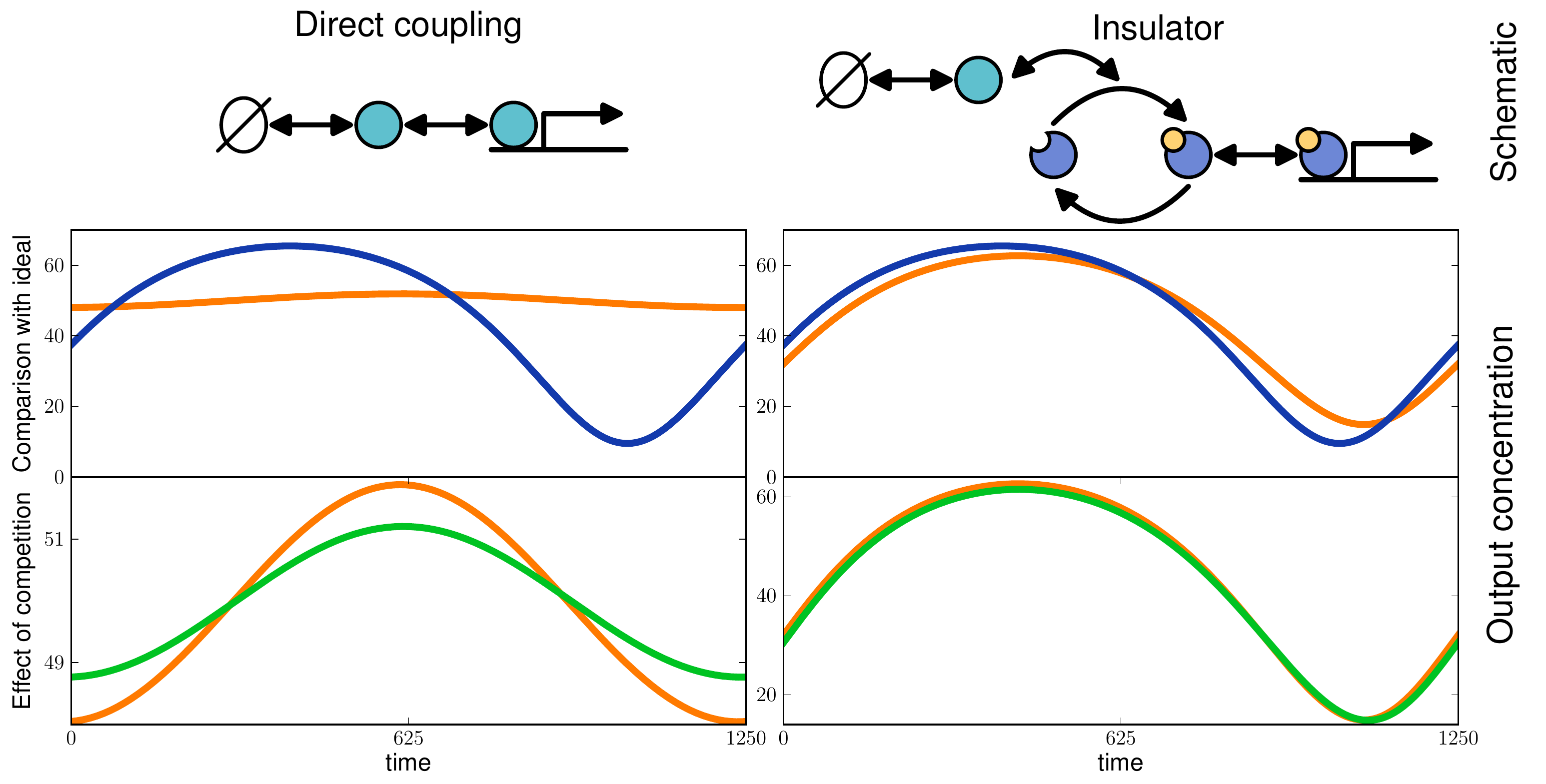}
\definecolor{bl}{HTML}{133AAC}
\definecolor{or}{HTML}{FF7A00}
\definecolor{gr}{HTML}{00C322}
\caption{
Retroactivity effects lead to signal distortion, and attenuation of output signals when additional targets are added.
Comparison of retroactivity effects on a signaling system with a direct coupling (DC) architecture (\textit{left}) and one with an insulator, represented by a phosphorylation/dephosphorylation cycle (\textit{right}).
Top row, a cartoon schematic of the signaling system. In the DC system \eqref{DC reactions}, the input binds directly to the target. With an insulator \eqref{PD reactions}, the input drives phosphorylation of an intermediate signaling molecule, whose phosphorylated form binds to the target.
Middle row, illustration of {\dn}. The ``ideal'' output signal (\textit{\color{bl}{blue}}), see \eqref{ideal ODEs}, with retroactivity effects neglected, is plotted against the output for each system with nonlinear dynamics (\textit{\color{or}{orange}}), given by \eqref{DC ODEs} for the DC system and \eqref{PD ODEs} for the insulator.
Bottom row, illustration of {\ce}. The output signal in a system with a single target (\textit{\color{or}{orange}}) is compared with the output signal when multiple targets are present (\textit{\color{gr}{green}}). Note the greatly reduced amplitude of variation of the output in the DC system.
Plots of the output signals in each system are shown in the steady state, over a single period of $k(t)$. This plot was made using the parameters $k(t)=0.01\left(1+\sin{\left(0.005\,t\right)}\right)$, $\delta=0.01$, $\alpha_1=\beta_1=0.01$, $\alpha_2=\beta_2=k_1=k_2=10$, $k_{\rm on}=k_{\rm off}=10$, $p_{\rm tot}=100$, and $X_{\rm tot}=800$, $Y_{\rm tot}=800$ for the insulator. Parameters specifying the interaction with the new promoter p$^\prime$ in the perturbed system are $k^\prime_{\rm on}=k^\prime_{\rm off}=10$, and $p^\prime_{\rm tot}=60$. 
\label{fig:output}} 
\end{figure}

As shown in Fig.~\ref{fig:output}, typical performance of the simple direct
coupling system defined by \eqref{DC ODEs} is poor, assuming, as
in~\citep{DelVecchio:2008gy},
that we test the system with a simple sinusoidally varying production rate 
\begin{equation} \label{k(t)}
k(t)=k\left(1+\sin{\omega t}\right),
\end{equation}
whose frequency $\omega$ is similar in magnitude to $k$ and $\delta$.
Oscillation of the output signal in response to the time-varying input is
strongly damped relative to the ideal. The output is also sensitive to other
targets connected in parallel; as the total load increases the output signal
is noticeably damped in both the transient and steady state. Here the flux of
Z into and out of the system is too slow to drive large changes in the output
C as the rate of production $k(t)$ varies.

As suggested in~\citep{DelVecchio:2008gy}, the retroactivity effects in this
system can be significantly ameliorated by using an intermediate
signal processing system, specifically one
based on a phosphoryl\-ation/de\-phosphorylation (PD)
``futile cycle,'' between the input and output.
Such systems appear often in signaling pathways that mediate gene expression
responses to the environment~\citep{alonbook}.
In this system, the input signal Z plays
the role of a kinase, facilitating the phosphorylation of a protein X. The
phosphorylated version of the protein X$^*$ then binds to the target p to
transmit the signal. Dephosphorylation of X$^*$ is driven by a phosphatase
Y. Assuming a two-step model of the phosphorylation/dephosphorylation
reactions, the full set of reactions is 
\begin{align} \label{PD reactions}
\begin{aligned}
& \emptyset \overset{k(t)}{\rw} \mbox{Z} \overset{\delta}{\rw} \emptyset , \\
& \mbox{Z+X} \overset{\beta_1}{\underset{\beta_2}{\rightleftharpoons}} \mbox{C}_1 \overset{k_1}{\rw} \mbox{X}^*\mbox{+Z} , \\
& \mbox{Y+X}^* \overset{\alpha_1}{\underset{\alpha_2}{\rightleftharpoons}} \mbox{C}_2 \overset{k_2}{\rw} \mbox{X+Y} , \\
& \mbox{X}^*\mbox{+p} \overset{k_{\rm on}}{\underset{k_{\rm off}}{\rightleftharpoons}} \mbox{C} . 
\end{aligned}
\end{align}
The total protein concentrations $X_{\rm tot}$ and $Y_{\rm tot}$ are
fixed. The forward and reverse rates of the phosphorylation/dephosphorylation
reaction depend implicitly on the concentrations of phosphate donors and
acceptors, such as ATP and ADP. Metabolic processes ensure that these
concentrations are held far away from equilibrium, biasing the reaction rates
and driving the phosphorylation/dephosphorylation cycle out of equilibrium.
As routinely done in enzymatic biochemistry analysis, we have made the
simplifying assumption of setting the small rates of the reverse processes 
$\mbox{X}^*\mbox{+Z}\rw \mbox{C}_1$ and $\mbox{X+Y}\rw \mbox{C}_2$ to zero. The ODEs governing the dynamics of the
system are then
\begin{align} \label{PD ODEs}
\begin{aligned}
&\frac{d Z}{d t}    	= k(t)-\delta Z - \beta_1 Z \left(X_{\rm tot} - C_1 - C_2 - C\right) + \left(\beta_2 +k_1\right) C_1 , \\
&\frac{d C_1}{d t}	= \beta_1 Z \left(X_{\rm tot} - C_1 - C_2 - C\right) - \left(\beta_2 +k_1\right) C_1 , \\
&\frac{d C_2}{d t}	= \alpha_1 X^* \left(Y_{\rm tot} - C_2\right) - \left(\alpha_2 +k_2\right) C_2 , \\
&\frac{d X^*}{d t}	= k_1\, C_1 - \alpha_1 X^* \left(Y_{\rm tot} - C_2\right) + \alpha_2 C_2 - k_{\rm on} X^* \left(p_{\rm tot} - C\right) + k_{\rm off} \, C , \\
&\frac{d C}{d t}   	= k_{\rm on} X^* \left(p_{\rm tot} - C\right) - k_{\rm off} \, C .
\end{aligned}
\end{align} 
As shown in Fig.~\ref{fig:output}, 
for suitable choices of parameters
the output signal in the system
including the PD cycle is able to match the ideal output much more closely
than in the direct coupling system. The output signal is also much less
sensitive to changes in other targets connected in parallel than in the system
where the input couples directly to the promoter. One can think of this system
with the insulator as equivalent to the direct coupling system, but with
effective ``production'' and ``degradation'' rates 
\begin{equation} \label{effective flux}
k_{\rm eff}(t) = k_1 C_1 + \alpha_2 C_2, \qquad \delta_{\rm eff} = \alpha_1 \left(Y_{\rm tot} - C_2\right),
\end{equation}
which may be much larger than the original $k(t)$ and $\delta$, thus allowing 
the system with the insulator to adapt much more rapidly to varying input.

The fact that the PD cycle is driven out of equilibrium, therefore consuming
energy, is critical for its signal processing effectiveness.
Our focus will be on how the performance of the PD cycle as an insulator
depends upon its rate of energy consumption.

Our hypothesis is that
\emph{better insulation requires more energy consumption}.
In order formulate a more precise question, we need to find a proxy for energy
consumption in our simple model.

\subsubsection*{Energy use and insulation}

The free energy consumed in the PD cycle can be expressed in terms of the
change in the free energy of the system $\Delta G$ resulting from the
phosphorylation and subsequent dephosphorylation of a single molecule of
X. One can also measure the amount of ATP which is converted to ADP, which is
proportional to the current through the phosphorylation reaction
$\mbox{C}_1\rw \mbox{X}^*\mbox{+Z}$. In the steady state, since the phosphorylation and
dephosphorylation reactions are assumed to be irreversible and the total
concentration $X_{\rm tot}$ is fixed, the time averages of these two measures
are directly proportional. The average free energy consumed per unit time in
the steady state is then proportional to the average current
\begin{equation} \label{power}
J = \langle k_1 C_1 \rangle .
\end{equation}
Different choices of the parameters appearing in the
phosphorylation and dephosphorylation reactions, such as $k_1$, $k_2$, and
$X_{\rm tot}$, will lead to different rates of energy use and also different
levels of performance in terms of the {\ce} and {\dn}. We focus our attention
on the concentrations $X_{\rm tot}$ and $Y_{\rm tot}$ as tunable
parameters. While the reaction rates such as $k_1$ and $k_2$ depend upon the
details of the molecular structure and are harder to directly manipulate,
concentrations of stable molecules like X and Y can be experimentally
adjusted,
and hence the behavior of the PD cycle as a function of 
$X_{\rm tot}$ and $Y_{\rm tot}$ is of great practical interest.

\subsubsection*{Comparing different parameters in the insulator: Pareto optimality}

In measuring the overall quality of our signaling system, the relative
importance of faithful signal transmission, as measured by small {\dn}, and a
small {\ce}, will vary.  This means that quality is intrinsically a
multi-objective optimization problem, with competing objectives.
Rather than applying arbitrary weights to each quantity, we will instead
approach the problem of finding ideal parameters for the PD cycle from the
point of view of Pareto optimality, a standard approach to optimization
problems with multiple competing objectives which was originally introduced in
economics \citep{Shoval:2012ke}.
In this view, one seeks to determine the set of parameters of the system for
which any improvement in one of the objectives necessitates a sacrifice in one
of the others. Here, the competing objectives are the minimization of $\Dist$
and $\Comp$.

\emph{A Pareto optimal choice of the parameters is one for which there is no
  other choice of parameters which gives a smaller value of both $\Dist$ and
  $\Comp$.}
Pareto optimal choices, also called Pareto efficient points, give generically
optimum points with respect to arbitrary positive linear combinations $\alpha
\Dist + \beta \Comp$, thus eliminating the need to make an artificial choice
of weights.

\subsubsection*{An informal analysis}

A full mathematical analysis of the system~(\ref{PD ODEs}) of nonlinear ODE's
is difficult.  In biologically plausible parameter ranges, however, certain
simplifications allow one to develop intuition about its behavior.  We discuss
now this approximate analysis, in order to set the stage for, and to help
interpret the results of, our numerical computations with the full nonlinear
model.

We make the following \emph{ansatz: the variables $Z(t)$ and $X^*(t)$
evolve more slowly than $C_1(t)$, $C_2(t)$, and $C(t)$.}
Biochemically, this is justified because phosphorylation and dephosphorylation
reactions tend to occur on the time scale of seconds \cite{kholodenko2000,hornberg2005},
as do transcription factor promoter binding and unbinding events \cite{alonbook}, 
while protein production and decay takes place on the time scale of minutes \cite{alonbook}.
In addition, we analyze the behavior of the system under the assumption that
the total concentrations of enzyme and phosphatase, $X_{\rm tot}$ and $Y_{\rm tot}$,
are large. 
In terms of the constants appearing in~(\ref{PD ODEs}), we assume:
\begin{align}
\begin{aligned}
& K_1 = \frac{\beta_2 + k_1}{\beta_1} \gg X_{\rm tot} \gg 1, \\
& K_2 = \frac{\alpha_2 + k_2}{\alpha_1} \gg Y_{\rm tot} \gg 1, \\
& 1 \gg \left\{ k(t),\;\delta \right\} \ll k_{\rm on} \approx k_{\rm off}\,. \nonumber
\end{aligned}
\end{align}

Thus, in the time scale of $Z(t)$ and $X^*(t)$, we can make the quasi-steady
state (Michaelis-Menten) assumption that $C_1(t)$, $C_2(t)$, and $C(t)$ are at
equilibrium.  Setting the right-hand-sides of 
$\frac{d C_1}{d t}$, $\frac{d C_2}{d t}$, and $\frac{d C}{d t}$ to zero, and
substituting in the remaining two equations of~(\ref{PD ODEs}), we obtain the
following system:
\begin{eqnarray*}
\frac{d Z}{d t}   &\approx& k(t)-\delta Z, \\
\frac{d X^*}{d t} &\approx& k_1\, C_1 - k_2\, C_2  \,.
\end{eqnarray*}
The lack of additional terms in the equation for $Z(t)$ is a
consequence of the assumption that $K_1 \gg X_{\rm tot}$, which
amounts to a low binding affinity of Z to its target X (relative to the
concentration of the latter); this follows from a ``total''
quasi-steady state approximation as in \cite{ciliberto2007,segel_tqss_1996}. Observe
that such an approximation is not generally possible for the original
system \eqref{DC reactions}, and indeed this is the key reason for the
retroactivity effect \cite{DelVecchio:2008gy}. 

With the above assumptions, in the system with the insulator $Z(t)$ evolves 
approximately as in the ideal system~(\ref{ideal ODEs}).  In this quasi-steady
state approximation 
$C_1\approx (1/K_1)(X_{\rm tot} - C_1-C_2-C)Z \approx (1/K_1)X_{\rm tot}Z$
and
$C_2\approx (1/K_2)(Y_{\rm tot} - C_2)X^* \approx (1/K_2)Y_{\rm tot}X^*$,
and thus we have
\[
\frac{d X^*}{d t} \approx (k_1/K_1)X_{\rm tot}Z - (k_2/K_2)Y_{\rm tot}X^*\,.
\]
Finally, let us consider the effect of the following condition:
\begin{equation} \label{eq:main_constraint}
(k_1/K_1)X_{\rm tot} \approx (k_2/K_2)Y_{\rm tot} \gg 1 \,.
\end{equation}
If this condition is satisfied, then
$\frac{d X^*}{d t} \approx K (Z - X^*)$, with $K\gg1$, which means that
$X^*(t)\approx Z(t)$, and thus the equation for 
$\frac{d C}{d t}$ in~(\ref{PD ODEs}) reduces to that for the ideal
system~(\ref{ideal ODEs}).
In summary, if~(\ref{eq:main_constraint}) holds, we argue that the system with
the insulator will reproduce the behavior of the ideal system, instead of the
real system~(\ref{DC ODEs}).
Moreover, the energy consumption rate in~(\ref{power}) is proportional
to $k_1 C_1 \approx (k_1/K_1)X_{\rm tot}Z$, and hence will be large if
condition~(\ref{eq:main_constraint}) holds, which intuitively leads us to
expect high energy costs for insulation.

These informal arguments (or more formal ones based on singular perturbation
theory \cite{DelVecchio:2008gy}) justify the sufficiency, but not the
necessity, of condition~(\ref{eq:main_constraint}).  Our numerical results will
show that this condition is indeed satisfied for a wide range of parameters
that lead to good insulation.

\section{Results} \label{results}

\begin{figure}
\centering
\hspace*{-30pt}
\includegraphics[scale=1.14]{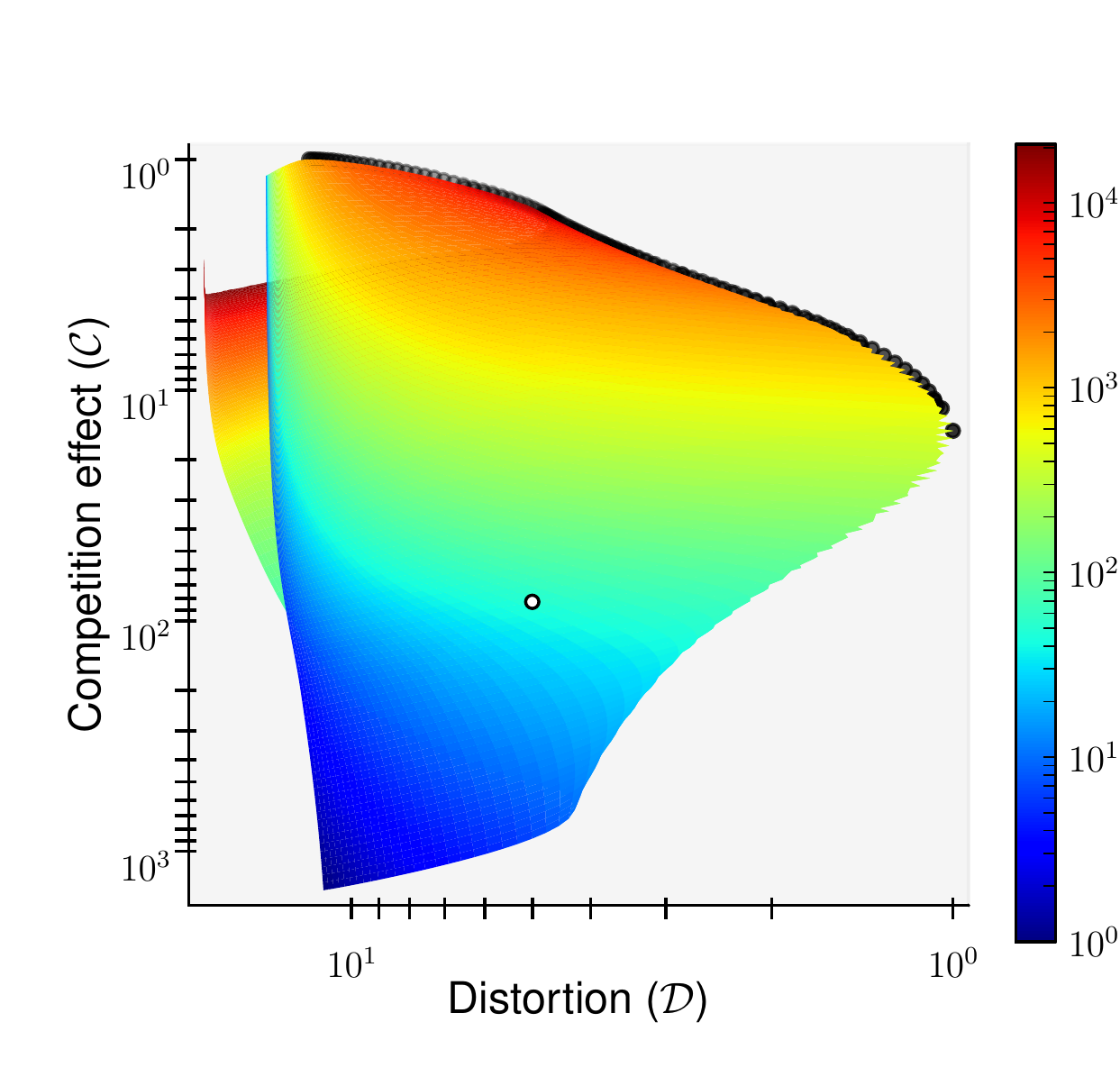}
\caption{
Performance of the insulator measured by the {\ce} $\Comp$ and {\dn} $\Dist$ of the output in the system with an insulator \eqref{PD ODEs}, tested over a range of $X_{\rm tot}$ and $Y_{\rm tot}$ varied independently from $10$ to $10000$ in logarithmic steps. For simplicity $\Comp$ and $\Dist$ are rescaled such that the smallest (best) values are equal to one. Points are colored according to the logarithm of the rate of the rate of energy consumption of the PD cycle, from blue (lowest) to red (highest). Pareto efficient parameter points are marked by black dots. Rates of energy consumption increases as one approaches the Pareto front; obtaining small values of the {\ce} is particularly costly. For comparison, $\Comp$ and $\Dist$ for the direct coupling system is marked by a white dot. See Section~\ref{results} for details. This plot was made using the parameters $k(t)=0.01\left(1+\sin{\left(0.005\, t\right)}\right)$, $\delta=0.01$, $\alpha_1=\beta_1=0.01$, $\alpha_2=\beta_2=k_1=k_2=10$, $k_{\rm on}=k_{\rm off}=10$, and $p_{\rm tot}=100$.
\label{fig:Pareto}} 
\end{figure}

\begin{figure}
\centering
\hspace*{-30pt}
\includegraphics[scale=0.80]{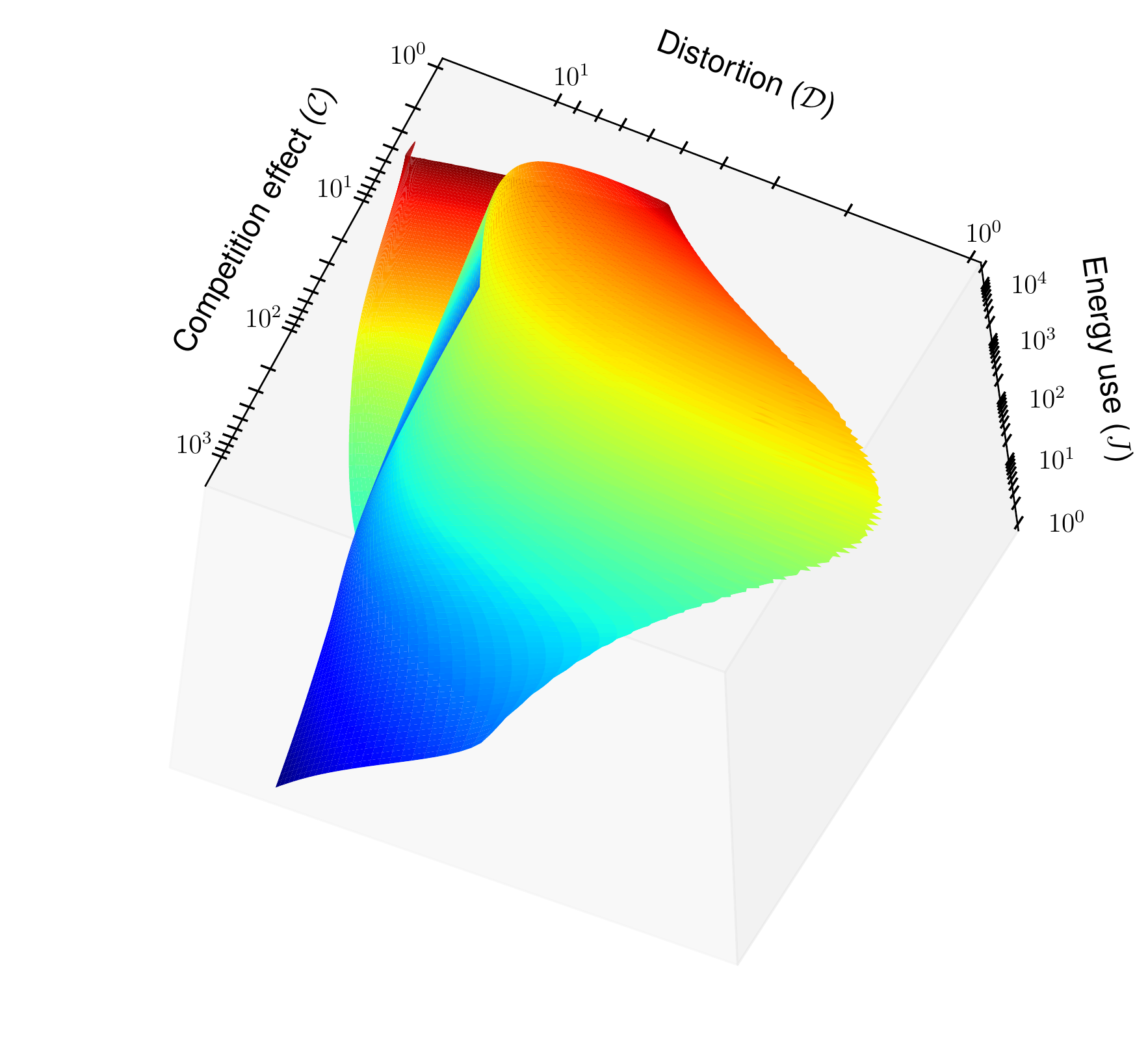}
\caption{
Three-dimensional plot of {\ce} $\Comp$ and {\dn} $\Dist$, along with the rate of energy consumption $J$, in the system including an insulator \eqref{PD ODEs}. Note the ``fold'' in the plot; it is possible for two different values of the parameters to yield the same measures of insulation $\Comp$ and $\Dist$, but with different rates of energy consumption.
\label{fig:Pareto3d}} 
\end{figure}

We have explored the performance of the insulating PD cycle over an extensive range 
of parameters to test our hypothesis that better insulation requires more energy
consumption. In Fig.~\ref{fig:Pareto} we show a plot of $\Comp$ and $\Dist$ for systems 
with a range of $X_{\rm tot}$ and $Y_{\rm tot}$, obtained by numerical integration of 
the differential equations \eqref{PD ODEs}(see also Fig.~\ref{fig:Pareto3d} for a 3d view). 
Pareto optimal choices of parameters on the tested parameter space are indicated by black 
points.

Superior performance of the insulator is clearly associated with higher rates of energy consumption,
as shown in the figure. Typically the rate of energy consumption increases as one approaches the set 
of Pareto optimal points, referred to as the Pareto front. Indeed, choices of parameters on or near the
Pareto front have some of the highest rates of energy expenditure. Conversely, the parameter choices 
which have the poorest performance also consume the least energy. As shown above, this phenomenon can 
be understood by noting that the energy consumption rate \eqref{power} will be large when the conditions
for optimal insulation \eqref{eq:main_constraint} are met.

Note that it is possible for two different choices of the parameters $X_{\rm tot}$ and $Y_{\rm tot}$ 
to yield the same measures of insulation $\Comp$ and $\Dist$, but with different rates of energy 
consumption. This results in a ``fold'' in the sheet in Fig.~\ref{fig:Pareto}, most clearly observed 
near $\Dist=10$ and $\Comp=4$. We see then that while better insulation generally requires larger 
amounts of energy consumption, it is not necessarily true that systems with high rates of energy 
consumption always make better insulators. See also the three-dimensional plot of $\Comp$,
$\Dist$, and $J$ shown in Fig.~\ref{fig:Pareto3d} for a clearer picture.

In addition to the general trend of increasing energy consumption as {\ce} or {\dn} decrease, we find 
that a strong local energy ``optimality'' property is satisfied. We observe numerically that any small 
change in the parameters $X_{\rm tot}$ and $Y_{\rm tot}$ which leads to a decrease in both the {\ce} 
and {\dn}, \emph{must} be accompanied by an increase in the rate of energy consumption, excluding 
jumps from one side of the ``fold'' to the other. This local property complements the global observation 
that Pareto optimal points are associated with the regions of parameter space with the highest rates 
of energy consumption.

\begin{figure}
\centering
\includegraphics[scale=0.50]{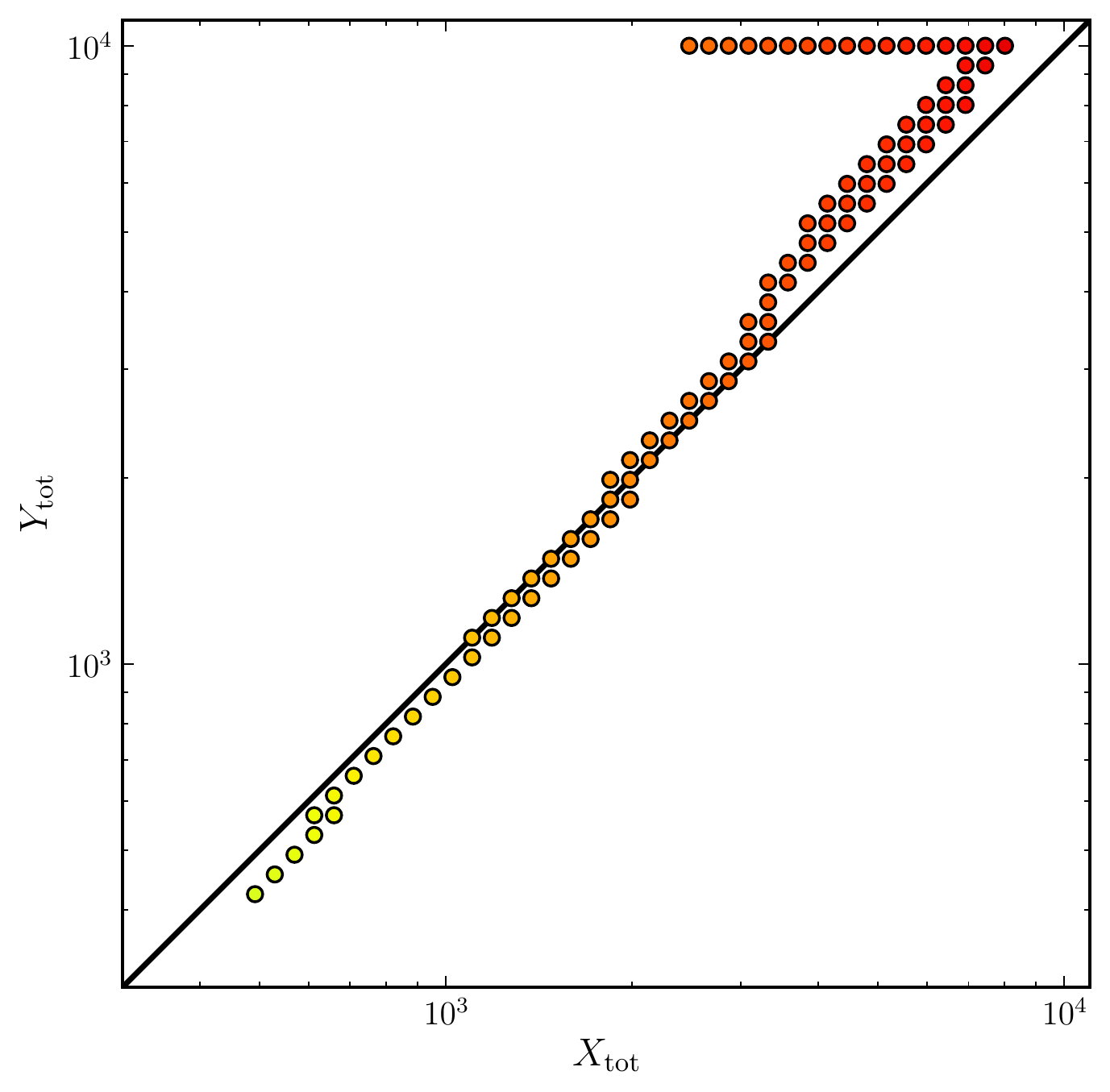}
\caption{
Pareto optimal points are typically those which strike a balance between the total concentrations of X and Y, as indicated in \eqref{eq:main_constraint}. Scatter plot of the Pareto optimal sets of parameters $X_{\rm tot}$ and $Y_{\rm tot}$ corresponding to those in Figs.~\ref{fig:Pareto}-\ref{fig:Pareto3d}. The background line shows $(k_1 / K_1) X_{\rm tot} = (k_2 / K_2) Y_{\rm tot}$ for comparison. Each point is colored according to the rate of energy consumption for that choice of parameters. Increases in either $X_{\rm tot}$ or $Y_{\rm tot}$ result in increased energy expenditure. Due to the limited range of parameters which could be tested, some Pareto optima lie along the boundaries of the parameter space (see the ``elbow'' in the scatter points at the top of the plot).
\label{fig:optima}}
\end{figure}

\begin{figure}
\centering
\includegraphics[scale=0.50]{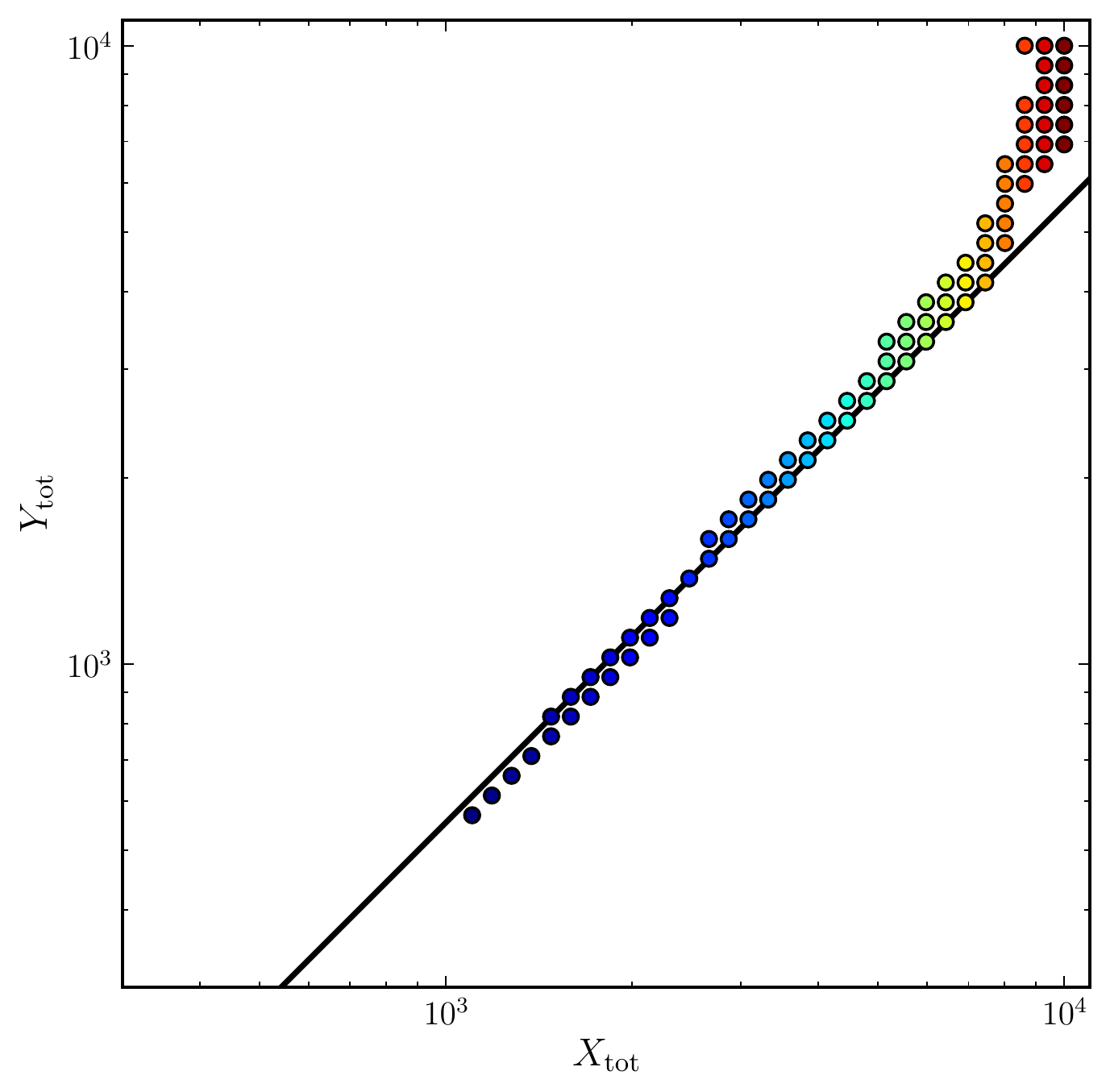}
\caption{
Scatter plot of the Pareto optimal sets of parameters $X_{\rm tot}$ and $Y_{\rm tot}$ for an insulator using the randomly shifted parameters. The background line shows $(k_1 / K_1) X_{\rm tot} = (k_2 / K_2) Y_{\rm tot}$ for comparison.  Parameters used are $k(t)=0.0137\left(1+\sin{\left(0.005\, t\right)}\right)$, $\delta=0.0188$, $\alpha_1=0.0107$, $\beta_1=0.0102$, $\alpha_2=5.31$, $\beta_2=16.42$, $k_1=k_2=10$, $k_{\rm on}=5.19$, $k_{\rm off}=12.49$, and $p_{\rm tot}=100$.  Each point is colored according to the rate of energy consumption for that choice of parameters (arbitrary scale, different from that of Figs.~\ref{fig:Pareto}-\ref{fig:optima}).
\label{fig:optimarand}}
\end{figure}

While we find Pareto optimal choices of the concentrations $X_{\rm tot}$ and $Y_{\rm tot}$ 
span several orders of magnitude, the \emph{ratio} of $X_{\rm tot}$ to $Y_{\rm tot}$ is close to 
unity for nearly all Pareto optima (see Fig.~\ref{fig:optima}). 
A small number Pareto optimal points are found with very different total 
concentrations of X and Y, but these points appear to be due to boundary effects from the sampling 
of a finite region of the parameter space. Indeed, we have argued that the insulator should perform
best when \eqref{eq:main_constraint} is satisfied. For the choice of parameters considered here, this gives 
$X_{\rm tot}/Y_{\rm tot} = (k_2 K_1)/(k_1 K_2) = 1$. Tests with randomized parameters confirm that 
\eqref{eq:main_constraint} gives a good estimate of the relationship between $X_{\rm tot}$ and 
$Y_{\rm tot}$ for Pareto optimal points (see Fig.~\ref{fig:optimarand} for an example). 

We also observe that there is a lower bound on the concentration of $X_{\rm tot}$ and $Y_{\rm tot}$ 
for optimal insulation. Though we tested ranges of concentrations from $10$ to $10000$, the first 
optimal points only appear when the concentrations are around $500$, several times larger than the 
concentration of the target p and much, much larger than the concentration of Z. Interestingly, the 
insulator consumes less energy for these first Pareto optimal parameter choices than at higher 
concentrations, and achieves the best measures of {\dn} with relatively low {\ce} as well. This 
suggests that smaller concentrations may be generically favored, particularly when energy constraints 
are important. 

We conclude that the specification of Pareto optimality places few constraints on the absolute 
concentrations $X_{\rm tot}$ and $Y_{\rm tot}$ in the model, save for a finite lower bound, 
but the performance of the insulator depends strongly on the ratio of the two concentrations. 
This observation connects with the work of Gutenkunst et al. \citep{sethna}, who noted ``sloppy'' 
parameter sensitivity for many variables in systems biology models, excepting some ``stiff'' 
combinations of variables which determine a model's behavior.

For comparison, we indicate the values of $\Dist$ and $\Comp$ of the simpler direct coupling 
architecture, with no insulator, by a white dot in Fig.~\ref{fig:Pareto}. While many choices of 
parameters for the insulating PD cycle, including most Pareto optimal points, lead to improvements 
in the {\dn} relative to that of the direct coupling system, the most dramatic improvement is in 
fact in the {\ce}. Roughly $85\%$ of the parameter values tested for the insulator have a lower 
value of the {\ce} than that found for the direct coupling system. This suggests that insulating PD 
cycles may be functionally favored over simple direct binding interactions particularly when there 
is strong pressure for stable output to multiple downstream systems.

Finally, we note that the analysis performed here has not factored in the potential 
metabolic costs of production for X and Y. Such costs would depend on the structure of these
components, as well as their rates of production and degradation, which we have not addressed 
and which may be difficult to estimate in great generality. However, as the rate of energy
consumption increases with increasing $X_{\rm tot}$ and $Y_{\rm tot}$ (see Fig.~\ref{fig:optima}), 
we would expect to find similar qualitative results regarding rates of energy consumption
even when factoring in production costs.

\section{Discussion}

A very common motif in cell signaling is that in which a
substrate is ultimately converted into a product, in an ``activation''
reaction triggered or facilitated by an enzyme, and, conversely, the product
is transformed back (or ``deactivated'') into the original substrate, helped
on by the action of a second enzyme.
This type of reaction, often called a futile, substrate, or enzymatic
cycle, appears in many signaling pathways:
GTPase cycles \cite{Donovan}, 
bacterial two-component systems and phosphorelays \cite{groisman, grossman}
actin treadmilling \cite{chen},
as well as glucose mobilization \cite{karp},
metabolic control \cite{stryer},
cell division and apoptosis \cite{sulis},
and cell-cycle checkpoint control \cite{lew}.
See~\cite{Samoilov} for many more references and discussion.

In this work we explored the connection between the ability of energy consuming enzymatic 
“futile cycles” to insulate biochemical signaling pathways from impedance and competition 
effects, and their rate of energy consumption. Our hypothesis was that better insulation
requires more energy consumption. We tested this hypothesis through the computational 
analysis of a simplified physical model of covalent cycles, using two innovative measures 
of insulation, referred to as {\ce} and {\dn}, as well as a new way to characterize optimal 
insulation through the balancing of these two measures in a Pareto sense. Our results indicate 
that indeed better insulation requires more energy.

Testing a wide range of parameters, we identified Pareto optimal choices which represent
the best possible ways to compromise two competing objectives: the minimization of {\dn} 
and of the {\ce}. The Pareto optimal points share several interesting features. First, 
they consume large amounts of energy, consistent with our hypothesis that better insulation
requires greater energy consumption. Second, the total substrate and phosphatase concentrations
$X_{\rm tot}$ and $Y_{\rm tot}$ typically satisfy \eqref{effective flux} (in natural units,
this implies $X_{\rm tot} \sim Y_{\rm tot}$). 
There is also a minimum concentration required to achieve a Pareto optimal 
solution; arbitrarily low concentrations do not yield optimal solutions. 
Interestingly, insulators with Pareto optimal choices of parameters close 
to the minimum concentration also expend the least amount of energy, compared to other 
parameter choices on the Pareto front, and have the least {\dn} while still achieving 
small {\ce}. This suggests that these points near the minimum concentration might be 
generically favored, particularly when energy constraints are important.

Many reasons have been proposed for the existence of futile cycles in nature,
such as signal amplification, increased sensitivity, and ``analog to digital''
conversion of help in decision-making.
An alternative, or at least complementary, possible
explanation~\citep{DelVecchio:2008gy} lies in the capabilities of such cycles
to provide insulation, thus enabling a ``plug and play'' interconnection
architecture that might facilitate evolution.
Our results suggest that better insulation requires a higher energy cost, so
that a delicate balance may exist between, on the one hand, the ease of
adaptation through creation of new behaviors by adding targets to existing
pathways, and on the other hand, the metabolic costs necessarily incurred in
not affecting the behavior of existing processes.


\end{document}